\documentstyle[12pt]{article}
\newcommand\be{\begin{equation}}
\newcommand\ee{\end{equation}}
\newcommand\bea{\begin{eqnarray}}
\newcommand\eea{\end{eqnarray}}
\newcommand\ket[1]{|#1\rangle}
\newcommand\bra[1]{\langle #1|}

\newcommand{\fatalpha}{{\bf \alpha \kern -0.44em \alpha}}
\newcommand{\fatsigma}{{\bf \sigma \kern -0.54em \sigma}}
\newcommand{\tpchi}{{\bf \chi \kern -0.35em \chi}}
\newcommand{\llambda}{{\bf \lambda \kern -0.45em \lambda}}
\newcommand\tr{\mbox{ Tr }}

\bibliography{plain}
\title{\bf  Quantum tomography  with wavelet transform in Banach space on Homogeneous space}
\vspace{20mm}
\author{ M. A. Jafarizadeh$^{a,b,c}$
 \thanks{E-mail:jafarizadeh@tabrizu.ac.ir},
M.Mirzaee$^{a,b}$
\thanks{E-mail:mirzaee@tabrizu.ac.ir},M.Rezaee$^{a,b}$
\thanks{E-mail:karamaty@tabrizu.ac.ir}
\\
\\
$^a${\small Department of Theoretical Physics and Astrophysics,
Tabriz University, Tabriz 51664, Iran.} \\ $^b${\small Institute for
Studies in Theoretical Physics and Mathematics, Tehran 19395-1795,
Iran.} \\ $^c${\small Research Institute for Fundamental Sciences,
Tabriz 51664, Iran.}} \pagebreak 
\begin{document}
\maketitle
\vspace{15mm}
\begin{abstract}
The intimate connection between the Banach space wavelet
reconstruction method on homogeneous spaces with both singular and
nonsingular vacuum vectors, and some of well known quantum
tomographies, such as: Moyal-representation for a spin, discrete
phase space tomography, tomography of a free particle, Homodyne
tomography, phase space tomography and SU(1,1) tomography is
explained. Also both the atomic decomposition and banach frame
nature of these quantum tomographic examples is explained in
details. Finally the connection between the wavelet formalism on
Banach space and Q-function is discussed.
\end{abstract}
\vspace{70mm}
\section{INTRODUCTION}

The mathematical theory of wavelet Transform finds nowadays an
enormous success  in various fields of science and technology,
including treatment of large databases, data and image
compression, signal processing, telecommunication and many other
applications \cite{meyer90}. After the empirical discovery by
Morlet \cite{Morlet}, it was recognized from the very beginning by
Grossmann, Morlet, Paul  and daubechies\cite{daubechies92} that
wavelets are simply coherent states associated to affine group of
the line (dilations and
translations)\cite{mallat89,daubechies99}. Thus, immediately the
stage was set for a far reaching
generalization\cite{daubechies92,Tomo10}.  Unlike function which
form orthogonal bases for space, Morlet wavelets are not
orthogonal and form frames. Frames are the set of functions which
are not necessarily orthogonal and which are not linearly
independent. Actually, frames are a repeatable set of vectors in
Hilbert space which produces each vectors in  space with a
natural representation.

Recently another concept called atomic decomposition have played
a key role in further mathematical development of wavelet theory.
Indeed atomic decomposition for any space of function or
distribution aims at representating any element in the form of a
set of simple function which are called atoms\cite{CHIRISTENSEN}.
As far as  the Banach space is concerned,  Feichtinger-Grocheing
\cite{Feichtinger}  provided a general and very flexible way to
construct coherent atomic decompositions and Banach frames for
certain Banach spaces, called coorbit spaces.

  The concept of a quantum state represents one of
the most fundamental pillars of the paradigm of quantum theory.
Usually the quantum state is described either by state vector in
Hilbert space, or density operator or a phase space probability
density distribution (quasidistributions). The quantum states can
be determined completely from the appropriated experimentally data
by using the well known technic of quantum tomography or better
to say  tomographic transformation.

A general framework is already presented for the unification of
the Hilbert space wavelets transformation on the one hand, and
quasidistributions and  tomographic transformation associated with
a given pure quantum states on the other hand \cite{MAnko}. Here
in this manuscript we are trying to present the intimate
connection between the Banach space wavelet reconstruction  method
developed by Feichtinger-Grocheing \cite{kisil,Feichtinger}  and
some of well known quantum tomographies associated with mixed
states, such as: Moyal-representation for a spin \cite{Weigert},
discrete phase space tomography \cite{Miquel}, tomography of a
free particle \cite{Jafar3}, Homodyne tomography
\cite{Sch,Gefen,Cle,Ste}, phase space tomography
\cite{Jafar3,Jafar1,Jafar2}and SU(1,1) tomography
\cite{Ariano01}, all which can be represented by density
matrices. Since the density matrix can be presented through
Banach space in quantum Physics \cite{arnold}. Therefore, it is
natural to do quantum tomography of each density matrix by using
the wavelet transform and its inverse in Banach space on
Homogeneous space corresponding to the associated density matrix.
The quantum tomography used by this method for the mixed quantum
states is completely consistent with other commonly used methods.
Also both the atomic decomposition and banach frame nature of
these quantum tomographic examples is explained in details.

The paper is organized as follows:\\ In section-2 we define
wavelet transform and its inverse  on homogeneous spaces with
both singlur and nonsingular vacuum vectors. In section -3 we
obtain some typical quantum tomographic examples with nonsingular
vacuum vectors, such as: Moyal-representation for a spin, discrete
phase space tomography, then define its atomic decomposition and
Banach frame bounds. In section -4 we  obtain some typical
quantum tomographic examples with singular vacuum vectors, such
as: Homodyne tomography, phase space tomography, SU(1,1)
tomography and tomography of a free particle and define its atomic
decomposition and Banach frame bounds. Finally, the connection
between the wavelet formalism on Banach space and Q-function is
discussed. The paper is ended with a brief conclusion.
\section{ Wavelet transform, frame and atomic decomposition  in
Banach spaces on homogeneous space:}
 The following is a brief
recapitulation of some aspects of the theory of  wavelets, atomic
decomposition and Banach frame on homogeneous space. We only
mention those concepts that will be needed in the sequel, a more
detailed treatment may be found  for example in
\cite{kisil,Feichtinger}. Let G be locally compact group with left
Haar measure $d\mu$ and H be a closed subgroup of G. Let $U$ be a
continuous representation of a group. The homogeneous space is
meant by $X=G/H$ .\\ Since U is not directly defined on G/H, it is
necessary to embed G/H in G. This can be realized by using the
canonical fiber bundle structure of G with projection
$\Pi:G\longrightarrow X$.  Let $\sigma:X\longrightarrow G$ be a
borel section of this fiber bundle i.e., $\Pi \circ \sigma(x)=x$
for all $x\in X$.

We could define a representation for homogeneous space $X \times
X$ in the space $\cal L(B)$ of bounded linear operators ${\cal
B\rightarrow B}$: \be T: X\times X\rightarrow {\cal L(L(B))}:
\hat{O} \rightarrow U(x_{1}) \hat{O} U(x_{2}^{-1}), \ee where if
$x_{1}$ is equal to $x_{2}$,  the representation is called adjoint
representation, and, if $x_{2}$  is equal to identity operator,
the representation is called left representation of homogeneous
space.

 Let $\cal L(B)$ be   the space of bounded linear operator
 ${\cal B}\rightarrow {\cal B}$ in Banach space.
  We will say that $b_{0}\in {\cal B}$ is a vacuum vector if for all $h\in
H$ then $U(h)b_{0}=\chi(h) b_{0}$ and also the set of vectors
$b_{x}=U(x) b_{0}$ forms a family of coherent states,  if there
exists a continuous non-zero linear functional $l_{0}\in {\cal
B}^{\ast}$ ( called test functional ) and a vector $b_{0}\in {\cal
B}$ ( called vacuum vector) such that
\begin{equation}\label{wave1}
C(b_{0},b^{'}_{0})=\int_{X}<T(x^{-1})b_{0},l_{0}><T(x)b^{'}_{0},l^{'}_{0}>
d\mu(x),
\end{equation}
is non-zero and finite, which is known as the admissibility
relation.\\If the subgroup $H$ is non-trivial, one does not need
to know wavelet transform on the whole group G, but it should be
defined on only the homogeneous space $G/H$, then the reduced
wavelet transform $\cal W$ to a homogeneous space of function F(X)
is defined by a representation $U$ of G on ${\cal B}$, a vacuum
vector $b_{0}\in {\cal B}$ and a test functional $l_{0}\in {\cal
B}^{\ast}$ such that\cite{kisil}
\begin{equation}\label{equ1}
{\cal W}: {\cal B}\rightarrow F(X): \hat{O} \rightarrow
\hat{O}(x)=[{\cal
W}\hat{O}](x)=<U(x^{-1})\hat{O},l_{0}>=<\hat{O},\pi^{\ast}(x)l_{0}>
\quad\forall x\in X.
\end{equation}
The inverse wavelet transform $\cal M$ from F(X) to ${\cal B}$ is
given by the formula:
\begin{equation}\label{equ2}
{\cal M}:F(X)\rightarrow {\cal B}:\hat{O}(x)\rightarrow {\cal
M}[\hat{O}]
=\int_{X}\hat{O}(x)b_{x}d\mu(x)=\int_{X}\hat{O}(x)U(x)b_{0}d\mu(x).
\end{equation}
The operator $P={\cal MW}: {\cal B}\longmapsto {\cal B}$ is a
projection of {\cal B} into its linear subspace in  which $b_{0}$
is cyclic (i.e., the set $\{T(x)b_{0}|x\in X\}$ span Banach space
$\cal B$), and ${\cal MW}(\hat{O})=P(\hat{O})$ in which  the
constant $P$ is equal to
$\frac{c(b_{0},b_{0}^{'})}{<b_{0},l_{0}^{'}>}$.  There are two
different cases which  correspond to different choices of vacuum
vector:

a) {\bf Non-singular cases:}\\ In this case, U is an irreducible
representation, then the inverse wavelet transform $\cal M$ is a
left inverse operator on ${\cal B}$ for the wavelet transform
$\cal W$ i.e., ${\cal MW}$=I for which  admissibility relation
(\ref{wave1}) holds.

b) {\bf Singular cases:}\\
 In this case  the representation U of G is neither
square-integrable  nor  square-integrable modulo  a subgroup H.
Therefore, the vacuum vector $b_{0}$ could not be selected within
the  original Banach space ${\cal B}$ (representation space of U
). Then, in the singular theorem, we assume that there is a
topological linear space $\hat{\cal B}$ with  ${\cal B}$ as its
subset such that:

1- ${\cal B}$ is dense in $\hat {\cal B}$ and representation U
could be uniquely extended to the continuous representation $\hat
U$ on $\hat {\cal B}$.

2- There exists $b_{0}\in {\hat {\cal B}}$ such that the following
relation holds for all $h \in H$ $$\hat{U}(h) b_{0}=\chi(h)
b_{0},\quad \quad \chi(h)\in C.$$

3- There exists a continuous non-zero linear functional $l_{0}\in
{\cal B}^{\ast} $ such that $U(h)^{\ast} l_{0}=\chi(h) l_{0}$

4- The following relation holds for a probe vector $p_{0} \in
{\cal B }$
\begin{equation}\label{equ4} C(b_{0}, p_{0})= <\int_{X}<U(x^{-1})
p_{0}, l_{0}> U(x) b_{0} d\mu(x), l_{0}>,
\end{equation}
where the integral converges in the weak topology of $\hat{{\cal
B}}$.

5- The composition ${\cal MW }: {\cal B}\rightarrow \hat{{\cal
B}}$ of the wavelet transform and the inverse wavelet transform
map ${\cal B}$ to ${\cal B}$.\\
 The choice
of probe vector is similar to regularization\cite{Wei}, which have
been used in our calculations. According to the theory of
distribution, the smoothness, regularity, and localization of a
temper distributions can be improved by a function of the Schwartz
class. Various regularizers can be used for numerical
computations.\\ A good example is the Gaussian distribution
:$$R_{\delta}(x)=exp(-\frac{x^{2}}{2\delta^{2}}),$$ where
$R_{\delta}$ is a regularizer which has properties \cite{Wei}
$$Lim_{\delta\longrightarrow \infty} R_{\delta}(x)=1,\quad
R_{\delta}(0)=1. $$ Frames can be seen as a generalization of
basis in Hilbert or Banach space\cite{atlanta}. Banach frames and
atomic decomposition are sequences that have basis-like properties
but which need not to be bases. Atomic decomposition has played a
key role in the recent development of wavelet theory.

 Now we define a  decomposition of a Banach space on homogeneous space as follow:

{\bf Definition of Coorbit space}: let ${\cal B}$ be a Banach
space and   ${\cal B}_{d}$ be an associated Banach space of
scalar-valued sequences indexed by $N=\{1,2,3,...\}$, and let
$\{y_{i}\}_{i\in N}\subset {\cal B}^{\ast}$ and $\{x_{i}\}_{i\in
N}\subset {\cal B}$ be given. The coorbit space is the  collecting
of all functions for which  wavelet transform is contained in
${\cal B}_{d}$. Similar to the  definition of coorbit space in
group, we can define coorbit spaces for X=G/H by \cite{coorbit}:
\begin{equation}\label{equ5}
M_{P}=\{\hat{O} \in {\cal B}: {\cal W} \hat{O}\in {\cal
B}_{d}\}\quad with \quad 1\leq d\leq\infty\quad and\quad norm
\quad ||\hat{O}||_{M_{P}}=||{\cal W} \hat{O}||_{{\cal B}_{d}}.
\end{equation}
{\bf Definition of atomic decomposition}: let $M_{P}$ be a coorbit
space and let ${\cal B}_{d}$ be an associated Banach space of
scalar-valued sequences indexed by $N=\{1,2,3,...\}$. Let
$\{y_{i}=\pi(\sigma^{-1}(x_{i}))l_{0}\}_{i\in N}\subset {\cal
B}^{\ast}$ and $\{\hat{O}_{i}=U(\sigma^{-1}(x_{i}))b_{0}\}_{i\in
N}\subset M_{P}$ be given. If \cite{coorbit}:

a ) $\{<\hat{O} , y_{i}>\} \in {\cal B}_{d}$ for each $\hat{O} \in
M_{P}$,

b ) The norms $\| \hat{O} \|_{M_{P}}$ and $\|\{ <\hat{O} ,
y_{i}>\}\|_{{\cal B}_{d}}$ are equivalent,

c ) $\hat{O} = \sum_{i=1}^{\infty}< O, y_{i}> x_{i}$ for each
$\hat{O} \in M_{P}$,\\ then $(\{y_{i}\}, \{x_{i}\})$ is an atomic
decomposition of X with respect to ${\cal B}_{d}$ and, if the norm
equivalence is given by:
\begin{equation}\label{dualsyms}
A\| \hat{O} \|_{M_{P}}\leq \|\{ <\hat{O} , y_{i}>\}\|_{{\cal
B}_{d}} \leq B\| \hat{O} \|_{M_{P}}, \end{equation} then A, B are
a choice of atomic bounds for $(\{y_{i}\}, \{x_{i}\})$. If $\bf i$
is a continuous index then $\sum_{i}\rightarrow\int d\mu(X)$.

{\bf Definition of Banach frame}: let $M_{P}$ be a coorbit space
and let ${\cal B}_{d}$ be an associated Banach space of
scalar-valued sequences indexed by $N=\{1,2,3,...\}$. Let
$\{y_{i}=\pi(\sigma^{-1}(x_{i}))l_{0}\}_{i\in N}\subset {\cal
B}^{\ast}$ and $\{\hat{O}_{i}=U(\sigma^{-1}(x_{i}))b_{0}\}_{i\in
N}\subset M_{P}$  and $S:{\cal B}_{d}\longrightarrow M_{P}$ be
given. If \cite{coorbit}

a ) $\{<\hat{O} ,y_{i}>\} \in {\cal B}_{d}$ for each $\hat{O} \in
M_{P}$,

b ) The norms $\| \hat{O} \|_{M_{P}}$ and $\|\{ <\hat{O} ,
y_{i}>\}\|_{{\cal B}_{d}}$ are equivalent. so that,$$A\| \hat{O}
\|_{M_{P}}\leq \|\{ <\hat{O} , y_{i}>\}\|_{{\cal B}_{d}} \leq B\|
\hat{O} \|_{M_{P}},$$

c ) S is bounded and linear, and $S\{<\hat{O} , y_{i}>\}=\hat{O}$
for each $\hat{O}\in M_{P}$.\\Then $(\{y_{i}\},S)$ is a Banach
frame for $M_{P}$ with respect to ${\cal B}_{d}$. The mapping S is
a reconstruction operator. If the norm equivalence is given by
$A\| \hat{O} \|_{M_{P}}\leq \|\{ <\hat{O} , y_{i}>\}\|_{{\cal
B}_{d}} \leq B\| \hat{O} \|_{M_{P}}$, then A, B are a choice of
frame bounds for $(\{y_{i}\}, S)$.\\ It is a remarkable fact that
the admissibility condition is a relation analogous to frame.
Again if {\bf i} is a continuous index then
$\sum_{i}\rightarrow\int d\mu(X)$.
\section{Quantum
tomography with wavelet transform on homogeneous space
(non-singular case)} \setcounter{equation}{0}
\subsection{ Moyal-type representations for a spin}
 In Moyal's formulation of quantum mechanics, a quantum spin $\bf s$ is described
in terms of {\em continuous symbols} i.e., by smooth functions on
a two-dimensional sphere. Such prescriptions to associate
operators with Wigner functions, $P$- or $Q$-symbols, are
conveniently expressed in terms of operator kernels satisfying the
Stratonovich-Weyl postulates. Similar to this approach, a {\em
discrete} Moyal formalism is defined on the basis of a modified
set of postulates\cite{Weigert}.
\begin{equation} \label{suchaniceform}
\hat\Delta_{\bf n}
            = \widehat U_{\bf n}  \hat\Delta_{\bf n_{z}}
                           \hat U_{\bf n}^\dagger,
\label{beforecontraction}
\end{equation}
where $\hat U_{\bf n}$ represents a rotation which maps the vector
${\bf n}_z$ to ${\bf n}$.\\By defining the  associated kernel as
\begin{eqnarray} \label{contcohpro}
    \hat{\Delta}_{\bf n}&=&\ket{s, {\bf n}}\bra{s, {\bf n}}
                               \equiv \ket{{\bf n}}\bra{{\bf n}}
                               \end{eqnarray}
\begin{eqnarray*} \hat{\Delta}^{\bf n}&=&\sum_{m=-s}^s
     \sum_{l=0}^{2s} \frac{2l+1}{2s+1}\left(\begin{array}{ccc}
s & l & s \\ s & 0 & s
\end{array}
\right)^{-1} \left(
\begin{array}{ccc}
s & l & s \\ m & 0 & m
\end{array}
\right)\ket{m,{\bf n}}\bra{m,{\bf n}}
        \end{eqnarray*}\label{PQkernels}
\begin{equation}=\sum_{m=-s}^s \Delta^{m}\ket{m,{\bf n}}\bra{m,{\bf
n}}.\end{equation} The reconstruction relation can be written as
\begin{equation}\label{eqmoyal}
\hat{O}=\frac{(2s+1)}{4\pi}\int_{S^{2}}dn Tr[\hat{O}
\hat{\Delta}_{\bf n} ] \hat{\Delta}^{\bf n}.
\end{equation}
In the wavelet notation, the Banach space is
$(2s+1)^{2}$-dimensional and group is $SU(2)$, the subgroup is
U(1) and measure is $d\mu(n)=\frac{2s+1}{4\pi} d(\bf n)$ and the
unitary irreducible representation of group is $U_{n}$ which is
the result of with adjoint representation on the any operators in
Banach space:
\begin{equation}
\hat{T}(n) \hat{O}=\hat{U_{n}} \hat{O} \hat{U}_{n}^{\dagger}.
\end{equation}
Then the wavelet transform in this Banach space with the test
functional, $$l_{0}(\hat{O})=Tr(\hat{O}\sum_{m}\Delta^{m}
\ket{m,n_{z}} \bra{m,n_{z}} ),$$ is given by:
\begin{equation}
{\cal W}:\hat{O} \longrightarrow \hat{O}(n)=<\hat{T(n)}^{\dagger}
\hat{O},l_{0}>=Tr(\hat{U_{n}}^{\dagger} \hat{O}
\hat{U_{n}}\sum_{m}\Delta^{m}\mid m,n_{z}><m,n_{z}\mid),
\end{equation}
then we have:\\ $$\hat{O}(n)=Tr(\hat{O}
\hat{U_{n}}\sum_{m}\Delta^{m}\mid m,n_{z}><m,n_{z}\mid
\hat{U_{n}}^{\dagger})=Tr(\hat{O}\hat{\Delta}^{n}).$$\\ If we
choose vacuum vector  $b_{0}=\mid s,n_{z}><s,n_{z}\mid$,  the
inverse wavelet transform ${\cal M}$ becomes left inverse operator
of the wavelet transform ${\cal W}$:
\begin{equation}
{\cal MW}=PI \Rightarrow {\cal M}:\hat{O}(n)\longrightarrow {\cal
M }(\hat{O})=\int<\hat{T}^{\dagger}(n)\hat{O},l_{0}>\hat{T(n)}
b_{0}
\end{equation}
$$=\int d\mu(n)
Tr(\hat{O}\hat{\Delta}^{n})\hat{U_{n}}\mid s,n_{z}><s,n_{z}\mid
\hat{U_{n}}^{\dagger}\Rightarrow
\hat{O}=\frac{1}{P}(\frac{2s+1}{4\pi}\int dn
Tr(\hat{O}\hat{\Delta^{n}})\hat{\Delta_{n}}).$$By using the
relations:
$$\frac{2s+1}{4\pi}\int_{{\cal S}^2} d{\bf n}
      \tr \left[ {\hat\Delta}_{\bf m} {\hat \Delta}^{\bf
          n}\right]  {\hat \Delta}_{\bf n}
      = {\hat \Delta}_{\bf m} \, , $$ and $$\tr \left[ {\hat\Delta}_{{\bf n}_z} {\hat \Delta}^{\bf n}
\right]
  = \sum_{l=0}^{2s} \frac{2l+1}{2s+1} P_l(\cos\theta)\, .
\label{tracial3}$$
 One can show that the constant on the left hand
side of (\ref{wave1}) is $C(b_{0}, b^{\prime}_{0})=2s+1$ and the
constant $P=\frac{C(b_{0}, b^{\prime}_{0})}{<b_{0}, l_{0}>}=1$,
and finally the reconstruction procedure of wavelet transform
(operating the combination of wavelet transform and its inverse
one, ${\cal MW}$ on the operator $\hat{O}$ ) leads to the
tomography relation (\ref{eqmoyal}). \\ By the same choice as
above for vacuum vectors and test functions, we can get the
atomic decomposition and Banach frame for this example. To do it,
we need further to choose  the set $\{ \hat{T}(n)l_{0}\}\subset
{\cal B}^{\ast}$ as the index sequence of functional which belongs
to dual Banach space, then we can show the following conditions:

a) $\{<\hat{O} , \hat{T}(n) l_{0}>\}=\{Tr(\hat{T}^{\dagger}(n)
\hat{O} )\}\in {\cal B}_{d}$ for each $\hat{O} \in M_{P}$,

b) The norms $||\hat{O}||_{M_{P}}$ and $||\{Tr(
\hat{T}^{\dagger}(n)\hat{O})\}||={[\int Tr(\hat{T}^{\dagger}(n)
\hat{O}) \overline{Tr(\hat{T}^{\dagger}(n) \hat{O} )} d\mu
(n)]}^{\frac{1}{2}}$ are equivalent  such  that they can satisfy
the inequality (\ref{dualsyms}) with the atomic bounds A=B=1,
providing that we use the the Hilbert-Schmidt norm for the
operator $\hat{O}$ and if we use the relation (\ref{eqmoyal}) we
have:

c) $\hat{O}=\int Tr(\hat{T}^{\dagger}(n) \hat{O} )\hat{T}(n) b_{0}
d\mu (n)$,\\ Therefore, $\{\hat{T}(n) b_{0},\hat{T}(n) l_{0}\}$ is
an atomic decomposition of $M_{P}$ of bounded operators acting on
representation space with respect  to ${\cal B}_{d}$ with atomic
bounds A=B=1.

Finally, by the same choice of vacuum vector, test functional and
index sequence of functional as in the atomic decomposition case,
yield the required conditions (a) and (b) for the existence of
Banach frame as the atomic decomposition one, and in order to have
the last condition for the existence of atomic decomposition, we
can define the reconstruction operator S as follows:

c) $S\{Tr( \hat{T^{\dagger}(n)} \hat{O} )\}=\int
Tr(\hat{T^{\dagger}(n)} \hat{O} )\hat{T(n)} d\mu (n)=\hat{O}$ for
each $\hat{O} \in M_{P}$,\\ It is straightforward to show that
the operator S as defined above is a linear bounded operator.
Therefore, $\{\hat{T(n)} l_{0}, S \}$ is Banach frame for $M_{p}$
with respect to ${\cal B}_{d}$ with frame bounds A=B=1.

\subsection{\bf Discrete
phase space tomography} In ref \cite{Miquel} formalism was
applied  to represent  the states and the evolution of a quantum
system in phase space in finite dimensional Hilbert space and,
finally, it was discussed how to perform direct measurement to
determine the wigner function. This approach was based on the use
of phase space point operator to define Wigner function. For
discrete systems we can define finite translation operators
$\hat{Q}$ and $\hat{V}$, which respectively generate finite
translation in position and momentum. The translation operator
$\hat{Q}$ generates cyclic shifts in the position basis and is
diagonal in momentum basis:
\begin{equation}\label{diswigner1}
\hat{Q}^{m}\mid n>=\mid n+m>,\quad\quad \hat{Q}^{m}\mid
k>=exp(-2\pi imk/N)\mid k>.
\end{equation}
Similarly, the operator $\hat{V}$ is a shift in the momentum basis
and is diagonal in position basis :
\begin{equation}\label{diswigner2}
\hat{V}^{m}\mid k>=\mid k+m>,\quad\quad \hat{V}^{m}\mid
n>=exp(2\pi imn/N)\mid n>.
\end{equation}
 Now by identifying the corresponding displacement operators, the
discrete analogue of the phase space translation operator is given
by:
\begin{equation}\label{diswigner4}
\hat{U}(q,p)=\hat{Q}^{q}\hat{V}^{p}exp(i\pi pq/N).
\end{equation}
 Here we can define the point operator as:
\begin{equation}\label{diswigner5}
\hat{A}(q,p)=\frac{1}{(2N)^{2}}\sum_{n,m=0}^{2N-1}\hat{U}(m,k)exp(-2\pi
i\frac{(kq-mp)}{2N} ),
\end{equation}
 or as:
\begin{equation}\label{diswigner6}
\hat{A}(\alpha)=\frac{1}{2N}\hat{Q}^{q}\hat{R}\hat{V}^{-p}exp(i\pi
pq/N ).
\end{equation}
That $\hat{R}$ is parity operator and it is worth noting that the
phase space point operators have been defined on a lattice with
$2N\times 2N$ points, but it has be shown that there are only
$N^{2}$ independent phase space point operators on the set
$G_{N}=\{\alpha=( q , p ); 0 \leq q, p \leq N-1\}$. The tomography
relation is given by:
\begin{equation}\label{diswigner7}
\hat{\rho}=1/N\sum_{\alpha \in G_{N} }Tr(\hat{\rho}
\hat{U}^{\dagger}(\alpha))\hat{U}(\alpha)=4N\sum_{\alpha \in G_{N}
}Tr(\hat{\rho} \hat{A}(\alpha))\hat{A}(\alpha).
\end{equation}
where  $W(\alpha)=Tr(\hat{A}(\alpha)\hat{\rho})$ is Wigner
function.\\  Now we try to obtain the tomography equation
(\ref{diswigner7}) via wavelets transform in Banach space.
Obviously the group, subgroup and representation are finite
Heisenberg, its center and $U(\alpha)$ respectively. Then the
wavelet transform with the test functional $$l_{0}(O)=Tr(O)\;\;
\mbox{for any operator O }$$ is given by
\begin{equation}
{\cal W}:{\cal B}\longmapsto F(\alpha):\hat{\rho} \longrightarrow
\hat{\rho}(\alpha)=<\hat{\rho}
,l_{\alpha}>=<\hat{U^{\dagger}}(\alpha) \hat{\rho}
,l_{0}>=Tr(\hat{U^{\dagger}}(\alpha)\hat{\rho} ).
\end{equation}
Since the representation is an irreducible representation, the
inverse wavelet transform ${\cal M}$ will be the left inverse
operator of wavelet transform ${\cal W}$:
\begin{equation}
{\cal M}:F(\alpha)\longmapsto {\cal B}:
\hat{\rho}(\alpha)\longrightarrow {\cal
M}[\hat{\rho}]=\hat{\rho}=\sum_{\alpha \in G_{N} }<\hat{\rho}
 ,l_{\alpha}>b_{\alpha}=\sum_{\alpha \in G_{N}
}<\hat{U^{\dagger}}(\alpha) \hat{\rho},l_{0}>\hat{U}(\alpha)b_{0},
\end{equation}
We  can obtain tomography relation (\ref{diswigner7}), for the
admissible $b_{0}=I/N$.  By the  same choice as above for vacuum
vector and test functions, we can get the atomic decomposition and
Banach frame for this example. To do it, we need just to choose
the $\{ \hat{U(\alpha)}l_{0}\}\subset {\cal B}^{\ast}\}$, then we
can show that:

a) $\{<\hat{\rho} , \hat{U(\alpha)}l_{0}>\}=\{Tr(\hat{\rho}
\hat{U^{\dagger}}(\alpha))\}\in {\cal B}_{d}$ for each
$\hat{\rho} \in M_{P}$,

b) The norms $||\hat{\rho}||_{M_{P}}$ and $||\{Tr(\hat{\rho}
\hat{U^{\dagger}}(\alpha))\}||$ are equivalent and in the sense
that they satisfy the inequality (\ref{dualsyms}) with the atomic
bounds A=B=1, provided that we use  the Hilbert-Schmidt norm for
the operator $\hat{O}$ and if we use the relation
(\ref{diswigner6}), we have,

c) $\hat{\rho}=\sum_{\alpha} Tr(\hat{\rho}
\hat{U^{\dagger}}(\alpha))\hat{U(\alpha)}b_{0}$,\\ then
$\{\hat{U(\alpha)} b_{0}, \hat{U(\alpha)}l_{0}\}$ is a linear
atomic decomposition of $M_{P}$ with respect to ${\cal B}_{d}$.

Finally by the same choice of vacuum vector, test functional and
index sequence of functional as in the atomic decomposition case,
we can show that the required conditions (a) and (b) for the
existence of Banach frame as the atomic decomposition one, and in
order to have the last condition for the existence of atomic
decomposition, we can define the reconstruction operator S as
follows

c) $S\{Tr(\hat{\rho} \hat{U^{\dagger}}(\alpha))\}=\sum_{\alpha}
Tr(\hat{\rho} \hat{U^{\dagger}}(\alpha))
\hat{U(\alpha)}=\hat{\rho}$ for each $\hat{\rho} \in M_{P}$,\\
then $\{\hat{U(\alpha)}l_{0},S \}$ is a Banach frame for $M_{P}$
with respect to ${\cal B}_{d}$ with frame bounds A=B=1.
\section{Quantum tomography with wavelet transform on
Homogeneous space singular case} \setcounter{equation}{0}
\subsection{\bf Homodyne Tomography}
The problem of measuring the density matrix $\hat\rho$ of
radiation has been extensively considered both experimentally and
theortically\cite{Tomo1}. Homodyne tomography is presently the
only method that can be used to achieve such measurement. This
method is based on the idea that the density matrix can be
evaluated in optical Homodyne experiments from the collection of
quadrature probability distribution for the radiation state.  As
shown in \cite{Tomo6}, the matrix can be obtained after
calculating the Wigner function as the inverse Radon transform of
such quadrature distributions \cite{Ariano974}. Quantum homodyne
tomography is used in quantum optic at the measurement of the
quantum state of light. In this case, we get \cite{Sch,Gefen,Cle}:
\begin{equation}\label{homo1}
\hat{\rho}=\int_{\cal C}\frac{d^{2}\alpha}{\pi} Tr[\hat{\rho}
\hat{U^{\dagger}}(\alpha)] \hat{U}(\alpha),
\end{equation}
where $\hat{U}(\alpha)=exp(\alpha a^{\dagger}-\alpha^{\ast} a )$
is a displacement operator. By Changing  polar variable
$\alpha=\frac{i}{2}ke^{i\phi}$ this formula becomes
\begin{equation}\label{homo2}
\hat{\rho}=\int_{0}^{2\pi}\frac{d\phi}{\pi}\int_{-\infty}^{\infty}\frac{dk\mid
k\mid}{4} Tr[\hat{\rho} e^{ikX_{\phi}}]e^{ikX_{\phi}},
\end{equation}
where  $X_{\phi}=\frac{(a^{\dagger}e^{i\phi}+a e^{-i\phi})}{2}$ is
field-quadrature operators that is measured by balance Homodyne
\cite{Ste}.\\ Now we try to obtain the tomography equation
(\ref{homo1}) via wavelets transform in Banach space. Obviously
the group is Heisenberg.  Since the representation of $H^{R}$
fails to be square-integrable, according to Stone-Von Neumann
\cite{Pak}, we can factor out the center $H^{R}$ and consider only
the factor space.\\ For the vacuum vector and test functional, we
need to choose the identity operator and $l_{0}(O)=Tr[O]$ for any
operator O, respectively. Then the wavelet formula is given by:
\begin{equation}\label{homo4}
{\cal W }:{\cal B}\mapsto F(\alpha)
:\hat{O}\mapsto\hat{\rho}(\alpha)=<\hat{\rho} ,l_{\alpha}>=<
\hat{\rho},\hat{U}(\alpha)l_{0}>=<\hat{\rho}
\hat{U}(\alpha)^{\dagger},l_{0}>=Tr(\hat{\rho}
\hat{U}(\alpha)^{\dagger}),
\end{equation}
 But above reference state is not admissible. Thus  according the
 singular cases,  we must select a probe vector $p_{0}\in {\cal B}$
 in which   equation (\ref{equ4}) is  non-zero and finite.
 In this case, the probe vector is selected
by:
\begin{equation}\label{homo5}
p_{0}=\int\mid\alpha><\alpha\mid e^{(\frac{-\mid \alpha
\mid^{2}}{\Delta})}\frac{d^{2}\alpha}{\pi},
\end{equation}
where $\Delta$ is non-zero and finite and $b_{0}\in \hat{B}$ is
identity. Since the representation is irreducible  and $C(b_{0},
p_{0} )=\Delta$, then the inverse wavelet transform in $\cal M$ is
a left inverse operator on ${\cal B}$ for the wavelet transform
$\cal W$:
\begin{equation}
{\cal M}{\cal W}=I\Rightarrow{\cal M}:F(\alpha)\mapsto
B:\hat{\rho}(\alpha)\mapsto {\cal M}[\hat{\rho}]={\cal M}{\cal
W}(\hat{\rho}),
\end{equation} then;
\begin{equation}\label{homo6}
\hat{\rho}=\int d\mu(\alpha)<\hat{\rho}
,l_{\alpha}>b_{\alpha}=\int d\mu(\alpha)Tr(\hat{\rho}
\hat{U^{\dagger}}(\alpha)) \hat{U}(\alpha)b_{0},
\end{equation}
where  d$\mu(\alpha)=\frac{d^{2}\alpha}{\pi}$ is the invariant
measure of the group of translation and group is unimodular. For
$b_{0}$ is equal to I,  the reconstruction procedure of wavelet
transform (\ref{homo6}) leads to the tomography relation
(\ref{homo1}).

In this relation $Tr(\hat{\rho} \hat{U^{\dagger}}(\alpha))$ is
Wigner characteristic function. We also can obtain another
quasidistribution characteristic functions with choosing different
representations. For example, for P-function
 characteristic function \cite{Glauber}, Q-function characteristic
function \cite{Glauber}, Husimi characteristic
function\cite{Husimi}, Standard-ordered characteristic function
\cite{Mehta} and Antistandard-ordered characteristic function
\cite{Kirkwood}, we need to choose the representations,
$\hat{U}_{an}(\alpha)=e^{\alpha
\hat{a}^{\dagger}}e^{-\alpha^{\ast}\hat{a}}$,
$\hat{U}_{n}(\alpha)=e^{-\alpha^{\ast}a}e^{\alpha a^{\dagger}}$,
$\hat{U}_{h}(\nu)=e^{-\nu^{\ast}b}e^{\nu b^{\dagger}}$
($\hat{b}=\mu\hat{a}+\nu\hat{a}^{\dagger}$ and
$\mu^{2}-\nu^{2}=1$),
$\hat{U}_{s}(\xi.\eta)=e^{i\xi\hat{q}}e^{i\eta\hat{p}}$ and
$\hat{U}_{as}(\xi.\eta)=e^{i\eta\hat{p}}e^{i\xi\hat{q}}$,
respectively.

For the complex Fourier transform of the displacement operator
$\hat{U}$ \cite{Jafar3}
\begin{equation}
\hat{\cal U}(\alpha)=\int
\frac{d^{2}\xi}{\pi}\hat{U}(\xi)exp(\alpha\xi^{\ast}-\alpha^{\ast}\xi),
\end{equation}
 the expansion of the  operator in terms of
the operator $\hat{\cal U}(\alpha)$ is given by
\begin{equation}
\hat{\rho}=\int\frac{d^{2}\alpha}{\pi}W(\alpha)\hat{\cal
U}(\alpha),
\end{equation}
where $W(\alpha)$ is Wigner function. Also by defining complex
Fourier transform for each above representation, we can  get its
tomography relation for each quasidisrbution. Now we will try to
obtain the atomic decomposition and Banach frame for this example.
Let $M_{P}$ be a Banach space and let ${\cal B}_{d}$ be an
associated Banach space of scalar-valued sequences and let $\{\{
\hat{U}(\alpha)l_{0}\}\subset {\cal B}^{\ast}\}$. Finally by the
same choice of vacuum vector, test functional and index sequence,
we can show that required conditions (a), (b) and (c) are
satisfied by atomic bounds A=B=1. Therefore,
$\{\hat{U}(\alpha)b_{0},\hat{U}(\alpha)l_{0}\}$ is a linear atomic
decomposition of $M_{p}$ with respect to ${\cal B}_{d}$.
Similarly, by using the relation (\ref{homo1}) and definition S,
$\{\hat{U}(\alpha)b_{0}, S \}$ is Banach frame for $M_{p}$ with
respect to ${\cal B}_{d}$ with frame bounds A=B=1. We can
generalize single mode Homodyne tomography to multimode state,
too.  In the wavelet notation, the irreducible representation is
$\hat{U}=\hat{U}_{0}\bigotimes \hat{U}_{1}\bigotimes...\bigotimes
\hat{U}_{m}$, which
$\hat{U}_{j}=exp(z_{j}\hat{a_{j}^{\dagger}}-z_{j}^{\ast}\hat{a_{j}})$,
and reduced wavelets formula with choose $b_{0}=\hat{I}\bigotimes
\hat{I} \bigotimes...\bigotimes \hat{I}$ is given by: $${\cal W
}:B\mapsto F(z_{0},z_{1},...,z_{m})
:\rho\mapsto\hat{\rho}(z_{0},z_{1},...,z_{m})$$
\begin{equation}
=<\hat{\rho},l_{z_{0},z_{1},...,z_{m}}>=<\hat{\rho} ,
\hat{U}(z_{1},z_{2},...,z_{m})l_{0}>=<\hat{\rho}
\hat{U}^{\dagger}(z_{0},z_{1},...,z_{m}),l_{0}>=tr(\hat{\rho}
\hat{U}^{\dagger}(z_{0},z_{1},...,z_{m})).
\end{equation}
But this reference state is not admissible. Thus according the
singular cases, we  must select a probe vector $p_{0}\in {\cal B}$
in which  that  equation (\ref{equ4}) is non-zero and finite. In
this case, the probe vector is selected by:
\begin{equation}
p_{0}=\int\mid z_{0},z_{1},...,z_{m}><z_{0},z_{1},...,z_{m}\mid
e^{(\frac{-\sum_{j=0}^{m}\mid z_{j}
\mid^{2}}{\Delta})}d\mu(z_{0},z_{1},...,z_{m}),
\end{equation}
where
\begin{equation}
\mid z_{0},z_{1},...,z_{m}>=\mid z_{0}>\otimes\mid
z_{1}>\otimes,...,\otimes\mid z_{m}>,
\end{equation}
and
\begin{equation}
d\mu(z_{0},z_{1},...,z_{m})=\frac{d^{2}z_{0}}{\pi}\frac{d^{2}z_{1}}{\pi}\cdot\cdot\cdot\frac{d^{2}z_{m}}{\pi},
\end{equation}
 where  $\Delta$ is non-zero and finite, and
  $b_{0}\in
\hat{B}$ is identity.  Since the representation is irreducible
 and $c(b_{0}, p_{0})=\Delta^{m+1}$,  the inverse wavelet
transform in $\cal M$ is a left inverse operator on ${\cal B}$ for
the wavelet transform $\cal W$: $${\cal M}{\cal
W}=I\Rightarrow{\cal M}:F(z_{0},z_{1},...,z_{m})\mapsto
B:\hat{\rho}(z_{0},z_{1},...,z_{m})\mapsto {\cal M}[\hat{\rho}]$$
\begin{equation}
={\cal M}{\cal W}(\hat{\rho})=\hat{\rho}=\int
d\mu(z_{0},z_{1},...,z_{m})<\hat{\rho},l_{z_{0},z_{1},...,z_{m}}>b_{z_{0},z_{1},...,z_{m}},
\end{equation}
Then;
\begin{equation}
\hat{\rho}=\int_{\cal C}\frac{d^{2}z_{0}}{\pi}\int_{\cal
C}\frac{d^{2}z_{1}}{\pi}\cdot\cdot\cdot\int_{\cal
C}\frac{d^{2}z_{m}}{\pi} Tr[\hat{\rho}
\hat{U}^{\dagger}(z_{0},z_{1},...,z_{m})]\hat{U}(z_{0},z_{1},...,z_{m}).
\end{equation}
The atomic decomposition and Banach frame is similar to one mode
Homodyne, and A , B are equal to identity.

\subsection{Phase Space Tomography \cite{Jafar3,Jafar1,Jafar2}:}
Any  marginal distribution is defined as the Fourier transform of
the characteristic function ${\cal W}(X,\mu,\nu)=\int dk
e^{-ikX}<e^{ik(\mu\hat{q}+\nu\hat{p})}>$. This marginal
distribution is related to the state of the quantum system which
is expressed in terms of its Wigner function $W(q,p)$, as follows
\begin{equation}\label{phase1}
{\cal W}(X,\mu,\nu)=\int dk
e^{-ik(X-\mu\hat{q}-\nu\hat{p})}W(q,p)\frac{dkdqdp}{(2\pi)^{2}}.
\end{equation}
It is possible to express the Wigner function in terms of the
marginal distribution of homodyne outcomes through the tomographic
formula.  An invariant form connecting directly the marginal
distribution ${\cal W}(X,\mu,\nu)$ and any operator was found
\begin{equation}\label{phase2}
\hat{\rho}=\int dX d\mu d\nu {\cal W}(X,\mu,\nu) \hat{K}_{\mu\nu},
\end{equation}
where the kernel operator has the form:
\begin{equation}
\hat{K}_{\mu\nu}=\frac{1}{2\pi}e^{iX}e^{i\mu\nu}e^{-i\nu\hat{p}}e^{-i\mu\hat{q}}.
\end{equation}
Now  we can try to obtain the tomography equation (\ref{phase2})
via wavelets transform in Banach space. Obviously the group is
Heisenberg in phase space. For the vacuum vector and test
functional we need to choose the identity operator and
$l_{0}(O)=Tr[O]$ for any operator O, respectively.  If we apply
the induced wavelet transform for representation
$\hat{U}(\mu,\nu)=e^{-i(\mu\hat{q}+\nu\hat{p})}$,  we have:
$${\cal W }:{\cal B}\mapsto F(\mu,\nu) :\hat{\rho}
\mapsto\hat{\rho}(\mu,\nu)=$$
\begin{equation}
<\hat{\rho} ,l_{(\mu,\nu)}> =<\hat{\rho} ,\hat{U}(\mu,\nu)l_{0}>
=<\hat{\rho} \hat{U}^{\dagger}(\mu,\nu),l_{0}>=Tr(\hat{\rho}
\hat{U}^{\dagger}(\mu,\nu)).
\end{equation}
The vacuum vector   $b_{0}=\hat{I}$ is not admissible, then we
choose a probe vector with the coherent state in the phase space
\cite{Tomo10} which is a translated Gaussian wave packet:
\begin{equation}
\eta_{\sigma(q,p)}(x)=(\pi^{-1/4})exp[-i(\frac{q}{2}-x)p]exp[-\frac{(x-q)^{2}}{2}]
\end{equation}
\begin{equation}
p_{0}=\int\mid\eta_{\sigma(q,p)}><\eta_{\sigma(q,p)}\mid
exp[\frac{-(q^2+p^2)}{\Delta}]dq dp,
\end{equation}
and the singularity condition gives $C(b_{0}, p_{0}) =\Delta$.\\
Since the representation is irreducible, the  inverse wavelet
transform $\cal M$ is a left inverse operator on {\cal B} for the
wavelet transform $\cal W$: $${\cal M}{\cal W}=I\Rightarrow{\cal
M}:F(\mu,\nu)\mapsto B:\hat{\rho}(\mu,\nu)\mapsto {\cal
M}[\hat{\rho}]={\cal M}{\cal W}(\hat{\rho})=\hat{\rho} $$
\begin{equation}
 \hat{\rho}=\int
d\mu(\mu,\nu)<\hat{\rho},l_{(\mu,\nu)}>b_{(\mu,\nu)}=\int d\mu
d\nu Tr[\hat{\rho}
\hat{U}^{\dagger}(\mu,\nu)]\hat{U}(\mu,\nu)b_{0},
\end{equation}
Then for $b_{0}=\hat{I}$, we have:
\begin{equation} \hat{\rho}=\int d\mu d\nu
Tr[\hat{\rho}\hat{R}^{\dagger}(\mu,\nu)]\hat{R}(\mu,\nu)=\int
d\mu\nu
Tr[\hat{\rho}e^{i(\mu\hat{q}+\nu\hat{p})}]e^{-i(\mu\hat{q}+\nu\hat{p})}.
\end{equation}
After simple calculation, we can obtain (\ref{phase2}). The atomic
decomposition and Banach frame are similar to one mode Homodyne,
and A,B are equal to identity.

\subsection{$SU(1,1)$ Tomography:}
The Lie algebra $su(1,1)$ of the $SU(1,1)$ group is spanned by the
operators $\hat{K}_{+},\hat{K}_{-},\hat{K}_{z}$. The Casimir
invariant operator that labels all the unitary irreducible
representations of the group is given by
$(\hat{K}_{z})^{2}-1/2(\hat{K}_{+}\hat{K}_{-}+\hat{K}_{-}\hat{K}_{+})=k(k+1)\hat{I}$,
where the eigenvalue  K is also called the Bargeman index. \\
Then the tomographic formula is given by:
\begin{equation}\label{SU1}
\hat{\rho}=\frac{1}{\pi}\int_{0}^{2\pi} d\phi \int_{0}^{\pi}
d\theta
 \tanh(\theta)Tr\bf[
\{(-1)^{\hat{K}_{z}}e^{\theta(e^{-i\phi}\hat{K}_{-}-e^{i\phi}\hat{K}_{+})},\hat{K}_{z}\}_{+}
\hat{\rho}]\times
\end{equation}
\begin{center}
$e^{i\theta/2(e^{-i\phi}\hat{K}_{+}+e^{i\phi}\hat{K}_{-})}\hat{K}_{z}e^{-i\theta/2(e^{-i\phi}\hat{K}_{+}+e^{i\phi}\hat{K}_{-})}.$
\end{center}
In the following  section, we will try to obtain the tomography
equation (\ref{SU1}) via wavelets transform in Banach space.
Obviously the group is $SU(1,1)$, and subgroup is U(1) with
reference state $b_{o}=I$. By choosing \be
\hat{\pi}(x)=\hat{u}^{\dagger}(x)\hat{K}_{z} \hat{u}(x)\ee,
\be\hat{U}(x)=\{(-1)^{\hat{K}_{z}}e^{\theta(e^{i\phi}\hat{K}_{+}-e^{-i\phi}\hat{K}_{-})},\hat{K}_{z}\}_{+}\ee,
 where
$\hat{u}(\theta,\phi)\equiv
e^{-i\theta/2(e^{-i\phi}\hat{K}_{+}+e^{i\phi}\hat{K}_{-})}$\cite{Ariano01},
the wavelet transform is given by:
\begin{equation}
{\cal W}: {\cal B} \rightarrow F(x):\hat{\rho} \rightarrow
\hat{\rho}(x)=[{\cal W}\hat{\rho}](x)=<\hat{U}(x^{-1})\hat{\rho}
 , l_{0}>=<\hat{\rho} , \pi^{\ast}(x)l_{0}>=Tr[\hat{U}^{\dagger}(x) \rho],
\end{equation}
and  inverse wavelet transform is given by
\begin{equation}
{\cal M}:F(x)\rightarrow {\cal B}:\hat{\rho}(x)\rightarrow {\cal
M}[\hat{\rho}(x)]
=\int_{x}\hat{\rho}(x)b_{x}d\mu(x)=\int_{X}\hat{\rho}(x)\pi(x)b_{0}d\mu(x),
\end{equation}
where $\hat{\pi}(x)$ is dual of $\hat{U}(x)$.  The reference
state is $b_{0}=I$ but this reference state is not admissible.
Thus according the
 singular cases,  we must select a probe vector $p_{0}\in {\cal B}$
 in which  equation (\ref{equ4}) is non-zero and finite.
 In this case, the probe vector is selected by

\begin{equation}
p_{0}=\sum_{r}b^{r}\mid r><r\mid,
\end{equation}
where this probe vector is similar to thermal states described by
the density operator $\rho_{T}$
\begin{equation}
\rho_{T}=\frac{1}{1+\widetilde{N}}\sum_{r}(\frac{\widetilde{N}}{1+\widetilde{N}})^{r}\mid
r><r\mid,
\end{equation}
where
$\widetilde{N}\equiv<\rho_{T}N>=\frac{1}{exp(\hbar\omega/KT)-1}$,
and $N=a^{\dagger}a$. In the high temperature this thermal state
is proportional with identity. Since the representation is
irreducible and $C(b_{0},p_{0})=\frac{1}{1-b}$,  the inverse
wavelet transform in $\cal M$ is a left inverse operator on ${\cal
B}$. Then the tomography formula for $SU(1,1)$ group is given by
the formula (\ref{SU1}).

 Now we will obtain atomic decomposition and Banach
frame for this example. Let $M_{P}$ be a coorbit space and let
${\cal B}_{d}$ be an associated Banach space of scalar-valued
sequences. Let $\{ \hat{\pi}(x)l_{0}\}\subset {\cal B}^{\ast}$,
then we can show that:

a) $\{<\hat{\rho}, \hat{\pi}(x)l_{0}>\}=\{Tr(\hat{\rho}
\hat{U}^{\dagger}(x))\}\in {\cal B}_{d}$ for each $f \in M_{P}$,

b) The norms $||\hat{\rho}||_{M_{P}}$ and $||\{Tr(\hat{\rho}
\hat{B}^{\dagger}(x))\}||$ are equivalent in the sense that they
satisfy the inequality (\ref{dualsyms}) with the atomic bounds
A=B=1, provided that we use the Hilbert-Schmidt norm for the
operator $\hat{\rho}$
\begin{equation}
||Tr(\hat{\rho} \hat{U}^{\dagger}(x))||^{2}=\int d\mu(x)
Tr(\hat{\rho} \hat{U}^{\dagger}(x)) \overline{Tr(\hat{\rho}
\hat{\pi}(x))},
\end{equation}
Since the dual couple $\hat{U}(x)$ and $\hat{\pi}(x)$ satisfy the
orthogonality relation \cite{Ariano01}:
$$\delta_{mk}\delta_{nl}=\int d\mu(x)
<m|B^{\dagger}(x)|n><l|C(x)|>, $$ then; $$||Tr(\hat{\rho}
\hat{U}^{\dagger}(x))||^{2}=\int d\mu(x)
{\rho}_{mn}U^{\ast}_{mn}(x){\rho}^{\ast}_{kl}\pi_{kl}(x)=||\hat{\rho}||^{2},
$$ and if we use the relation (\ref{SU1}), we have:

 c) $\hat{\rho}=\int d\mu(x) Tr(\hat{\rho} \hat{U}^{\dagger}(x))\hat{\pi}(x)$,

 Therefore,
$\{\hat{\pi}(x)b_{0},\hat{\pi}(x)l_{0}\}$ is an atomic
decomposition of $M_{P}$ with respect to ${\cal B}_{d}$ with
atomic bounds A=B=1. Similar to atomic decomposition,
$\{\hat{\pi}(x)l_{0},S \}$ is a Banach frame for coorbit space of
operators with respect to ${\cal B}_{d}$ with frame bounds A,B are
equal to identity.
\subsection{Tomography of a free particle}
Here  we will  consider the tomography of a free particle.  For
simplicity we suppose  a particle with unit mass and use
normalized unit $\hbar/2 =1$, so that the free Hamiltonian is
given by $\hat H_F = \hat p^2$. The  basis is constituted by the
set of operator  $\hat
R(x,\tau)=e^{-i\hat{p}^{2}\tau}\ket{x}\bra{x}e^{i\hat{p}^{2}\tau}$
\cite{Jafar3}; then, a generic free particle density operator can
be written as:
\begin{eqnarray}\label{free3}
\hat{\rho}= \int_{R}\int_{R} dx \:d\tau\: p(x,\tau) \: \hat
R(x,\tau),\label{free3}\;
\end{eqnarray}
where $p(x,\tau)= \hbox{Tr}[\hat\varrho\:\hat R(x,\tau)]$  is the
probability density of the particle to be at position $x$  at
time $\tau$.

Now  we try to obtain the tomography equation (\ref{free3}) via
wavelets transform in Banach space. Obviously the group is $\{\hat
{P},\hat{X},\hat{P}^{2},I\}$ and subgroup is $\{\hat{X},I\}$. The
relevant representation for this example is adjoint
representation:
   $$\hat{T}(x,\tau) \hat{\rho}=
\hat{U}(x,\tau) \hat{\rho} \hat{U}^{-1}(x,\tau) \quad\quad with
\;\; \hat{U}(x,\tau)=e^{-i\hat{P}^{2}\tau}\hat{D}(x).$$In this
representation, $\hat{D}(x)$ is translation operator, so that
$\hat{D}(x)\ket{0}=\ket{x}$, where $\ket{x}$ is eigenstate of
position operator and $\hat{P}$ is the momentum operator. On the
other hand if we define:  $$<
\hat{\rho},l_{0}>=l_{0}(\hat{\rho})=Tr(\hat{\rho}\mid 0><
0\mid).$$  the wavelet transform formula is given by: $${\cal W
}:{\cal B}\mapsto F(x,\tau) :
\hat{\rho}\mapsto\hat{\rho}(x,\tau)=$$ $$< \hat{\rho} ,
l_{(x,\tau)}> =<\rho ,\hat{T}(x,\tau)l_{0}>
=<\hat{T}^{\dagger}(x,\tau) \hat{\rho} ,l_{0}>=$$
\begin{equation}
tr( \hat{T}^{\dagger}(x,\tau) \hat{\rho} \mid 0><
0\mid)=Tr(\hat{U}(x,\tau)\mid 0><0\mid \hat{U}^{\dagger}(x,\tau)
\hat{\rho})=Tr(\hat{\rho}e^{-i\hat{P}^{2}\tau}\mid x><x\mid
e^{i\hat{P}^{2}\tau}).
\end{equation}
Also the inverse wavelet transform $\cal M$ associated with
wavelet transform $\cal W$ is:\\ $$ {\cal MW}=PI\Rightarrow {\cal
M}:F(x,\tau)\mapsto {\cal B}:\hat{\rho}(x,\tau)\mapsto {\cal
M}[\hat{\rho}]=$$
\begin{equation}
\int d\mu(x,\tau)< \hat{\rho},l_{(x,\tau)}>b_{(x,\tau)}=\int dx
d\tau Tr[\hat{\rho}e^{-i\hat{P}^{2}\tau}\mid x><x\mid
e^{i\hat{P}^{2}\tau}]\hat{T(x,\tau)}b_{0},
\end{equation}
The vacuum vector  is $b_{0}=|0><0|$, but this vacuum vector  is
not admissible. Thus according the
 singular cases,  we must select a probe vector $p_{0}\in {\cal B}$
 in which  equation (\ref{equ4}) is non-zero and finite.
 In this case, the probe vector is selected by
\begin{equation}
p_{0}=\mid D><D\mid,
\end{equation}
where $<D\mid p> = e^{-\frac{p^{2}}{D}}$. Its follows from
bi-orthogonality and from the following relations
\cite{Jafar3}(for $| j\rangle$, $j=p_{1},p_{2},p_{3},p_{4}$)
\begin{eqnarray}\label{free2}
 \int_{R}\int_{R} &dx \:d\tau\: & \langle p_{1}|\hat
R(x,\tau)|p_{2}\rangle \: \langle p_{3}|\hat
R(x,\tau)|p_{4}\rangle \nonumber \\  &=& \int_{R}\int_{R} dx
\:d\tau \: e^{-i \tau (p_{2}^2-p_{1}^2 + p_{3}^2 -p_{4}^2)}\:
\langle p_{1}|x \rangle \langle x|p_{2}\rangle\: \langle p_{3}|x
\rangle \langle x|p_{4}\rangle \nonumber \\  &=& \int_{R}\int_{R}
dx\: d\tau \: e^{-i \tau (p_{2}^2-p_{1}^2 + p_{3}^2 -p_{4}^2)}\:
e^{i x (p_{1} -p_{2} + p_{3} -p_{4} )}\: \nonumber \\  &=&
\delta(p_{1}-p_{3})\:\delta(p_{2}-p_{4}) \label{free2}\;.
\end{eqnarray}
we can show that the constant on left hand side of (\ref{equ4})
is $C(b_{0},p_{0})=D/2\sqrt{\pi}$ and  finally the reconstruction
procedure of wavelet transform leads to the tomography relation
(\ref{free3}). In order to obtain atomic decomposition and Banach
frame for this example, let
 $M_{P}$ be a coorbit  space and let ${\cal B}_{d}$ be an associated
Banach space of scalar-valued sequences and  $\{\{
\hat{T}(x,\tau)l_{0}\}\subset {\cal B}^{\ast}\}$. Finally by the
same choice of vacuum vector, test functional and index sequence,
we can show that the required conditions (a), (b) and (c) are
satisfied by atomic bounds A=B=1. Therefore,
$\{\hat{T}(x,\tau)b_{0},\hat{T}(x,\tau)l_{0}\}$ is a linear
atomic decomposition of $M_{p}$ with respect to ${\cal B}_{d}$.
Similarly, by using the relation (\ref{free3}) and definition S,
$\{\hat{T}(x,\tau)b_{0}, S \}$ is Banach frame for $M_{p}$ with
respect to ${\cal B}_{d}$ with frame bounds A=B=1.
\subsection{Wavelet transform and Q-function:}
Let $g \in {\cal L}^{2}(R)$ with $\parallel g \parallel=1$ and the
time-frequency translation of $g$ be:
\begin{equation}
g^{[x_{1},x_{2}]}(t)=e^{2\pi i t
x_{2}}g(t+x_{2})=U[x_{1},x_{2},0]g(t),
\end{equation}
where $U$ is the unitary irreducible representation of the
Heisenberg group $H^{R}$. To consider an arbitrary function $f\in
L^{2}(R)$, we can compute the following inner product for pure
state sampling \cite{miller}:
\begin{equation}
F(x_{1},x_{2})=<f,g^{[x_{1},x_{2}]}>,
\end{equation}
where $g^{[x_{1},x_{2}]}=U[x_{1},x_{2}] g(t)$ is a coherent
state. For the pure states, square of sampling is Q-function.\\
Now we will try to obtain Q-function via wavelet and  we  will
show that the wavelet transform in the Banach space is
Q-function. The group is Heisenberg and subgroup is identity and
representation is adjoint.
 Then the wavelet transform is  given by:$${{\cal W }:{\cal B}\mapsto
F(\alpha) :}$$
\begin{equation}
\hat{\rho}\mapsto\hat{\rho}(\alpha)=<\hat{\rho},l_{\alpha}>=<\hat{\rho}
, \hat{T}(\alpha)l_{0}>=< \hat{T}(\alpha)^{\dagger} \hat{\rho}
,l_{0}>.
\end{equation}
On the other hand if we choose:
\begin{equation}
<\hat{\rho} ,
l_{0}>=l_{0}(\hat{\rho})=Tr[\hat{\rho}\ket{0}\bra{0}],
\end{equation}
Then the wavelet transform for the adjoint representation is given
by: $${\cal W }:{\cal B}\mapsto F(\alpha) :\hat{\rho}
\mapsto\hat{\rho}(\alpha)=Tr\{\quad \hat{T}(g)^{\dagger}
\hat{\rho} \ket{0}\bra{0} \}$$
\begin{equation}
=Tr\{\quad \hat{U}(\alpha)^{\dagger}(\hat{\rho}) \hat{U}(\alpha)
\ket{0}\bra{0}\}=\bra{0} \hat{U}(\alpha)^{\dagger}(\hat{\rho})
\hat{U}(\alpha) \ket{0}= \bra{\alpha} \hat{\rho}
\ket{\alpha}=Q(\alpha).
\end{equation}
\section{Conclusion}\label{secConclusion}
In this paper     we have generalized wavelet transform and its
inverse for tomography of density operator in Banach space on
homogeneous space. Also we have explained  some examples of the
using the wavelet  formalism in quantum tomography on homogeneous
space and introduced frame and atomic decomposition for each of
them. We have also presented the connection between the wavelet
formalism on Banach space and Q-function.


\begin{thebibliography}{99}
\bibitem{meyer90} Y.~Meyer, Wavelets:  algorithms and applications (SIAM), Philadelphia (1993).
\bibitem{Morlet} A. Grossmann, J. Morlet and T. Paul, J. Math.
Phys 26 (1985) 2473-2479.
\bibitem{daubechies92} I.~Daubechies, Ten Lectures on
Wavelets. Philadelphia: Society for Industrial and Applied
\bibitem{mallat89} S. Mallat, IEEE Trans.\ Pattern Anal.\
Mach.\ Intel.  11  (1989)  674-693.
\bibitem{daubechies99} I.~Daubechies,  Orthonormal bases of compactly supported
wavelets, Commun. Pur Appl. Math. 41   (1988)  909-996.
\bibitem{Tomo10} S T. Ali, J-P Antoine, J-P.Gazeau:  Coherent States, Wavelets and their Generalizations Springer
(2000).
\bibitem{kisil} Vladimir V. kisil,   Wavelets in Applied and Pure Mathematics,
Lecture note 22 May(2003).
\bibitem{miller} W. Miller,  Topics in Hormonic Analysis With Applycations
To Radar and Sonar  Lecture note 23 October (2002).
\bibitem{CHIRISTENSEN} O. Christensen, C. Heil, Math. Nachr.
185 (1997) 33-47.
\bibitem{Feichtinger}  H.G. Feichtinger and K.H. Grochenig, J. Functional Anal, 86, No 2, (1989) 308-339.
\bibitem{MAnko} M. A. Man'ko, V. I. Man'ko, R. Vilela Mendes,  J.
of Physics A: Math. and Gen. 34 (2001) 8321-8332
\bibitem{Weigert} S. Heiss, S. Weigert: Discrete Moyal-type
representations for a spin. Phys. Rev. A 63 (2001) 012105.
\bibitem{Miquel}  C, Miquel, J, P. Paz, M. Saraceno, Phy. Rev A 65  (2002)
259 (1995) 147-211. 062309.
\bibitem{Jafar3} G. M. D'Ariano, S. Mancini, V. I. Manko, P. Tombesi, J. Opt. B: quantum and semiclassical
opt. 8  (1996)  1017.
\bibitem{Sch}   M. Paini, quantu-ph/0002078.
\bibitem{Gefen} G. M. D'Ariano,L. Maccone and M. G. A.  Paris, J. Phys. A: Math, Gen. 34  No. 1 (12 Janury
2001) 93-103.
\bibitem{Cle}  G. M. D'Ariano,  Advances in Physics, vol  39 (1990) 191.
\bibitem{Ste}  G.M. D'Ariano, L. Maccone, M. Paini, J. Opt. B: quantum semicalss. Opt. 5  (2003)  77.
\bibitem{Jafar1} T. J. Dunn, I .A Walmsley, S. Mukamel, Phys. Rew. Lett Vol 74 (1995) 884.
\bibitem{Jafar2} Mancini, Manko, V.I. Manko, P. Tombesi,
J.Phys.A: Math. Gen 34 (2001) 3461.
\bibitem{Ariano01} G. M. D'Ariano, E. De Vito and L. Maccone, Phys. Rev A
64 (2001) 033805.
\bibitem{arnold} E. B. Davis,  Quantum theory of open system,Academic Press
(1976).
\bibitem{Tomo1} U. Leonhardt, Measuring the quantum state of light (Cambridge University Press, Cambridge, England 1997)
\bibitem{Tomo6} K. Vogel and Risken, Phys. Rev. A 40 (1989) 2847.
\bibitem{peter} P. G. Gasazza , Advances in Computational Mathematics, special issue on frames, (2002).
\bibitem{coorbit} S. Dahlke, G. Steidl and  G. Teschke, Coorbit
spaces and Banach frames on Homogeneous spaces with application to
analizing function on sphers , ZeTeM Thecnical report 01-13,
(11/2001), To appear in: Adv. Comput. Math.ibid   Weighted Coorbit
spaces and Banach frames on homogeneous spaces, ZeTeM Thecnical
report 03-4, (2003), To appear in: Adv. Comput. Math.
\bibitem{Wei} G. W. Wei, Y. B. Zhao and Y. Xiang, Int. J. Numer. Math. Eng. 55 (2002) 913-946.
\bibitem{atlanta} O. Christensen, C. Heil, Math. Nachr. 185 (1997)
 33-47.
\bibitem{Ariano974} Giacomo mauro D'Ariano and Nicoletta Sterpi, J. modern
Opt. 44 (1997) 2227-2232.
\bibitem{Pak} A. A. Kirillov,  Elements of the theory of representation, Springer-Verlag, berlin (1976).
\bibitem{Glauber} R. J. Glauber, Phys. Rev. 130 (1963a) 2529.
\bibitem{Husimi} K. Husimi, Proc. Phys. Mat. Soc. Jpn, 22 (1940) 264-314.
\bibitem{Mehta} c. l. Mehta. J. Math. Phys. 5 (1940) 69.
\bibitem{Kirkwood} J. G. Kirkwood, Phys. Rev, 44 (1933) 31.
\end{thebibliography}
\end{document}